\documentclass{article}
\usepackage[english]{babel}

\begin{document}

% \title{Is a Two Composite Higgs Doublets Model the Low Energy Limit of a
% Natural Strong Electroweak Symmetry Breaking Sector ?}

\title{Two Composite Higgs Doublets: Is it the Low Energy Limit of a
 Natural Strong Electroweak Symmetry Breaking Sector ?}
\author{Alfonso R. Zerwekh\thanks{alfonsozerwekh@uach.cl}\\
Centro de Estudios Subat\'omicos and Instituto de F\'isica,\\ Facultad
de Ciencias,\\ Universidad Austral de Chile,\\ Casilla 567, Valdivia,Chile}

\date{}

\maketitle

\begin{abstract}
In this paper, we propose an effective model scheme that describes the
electroweak symmetry breaking sector by means of composite Higgs-like
scalars, following the ideas of Minimal Walking Technicolor (MWT).
We argue that, because of the general failure of Extended Technicolor
(ETC) to explain the mass of the top quark, it is necessary to introduce
two composite Higgs bosons: one of them originated by a MWT-ETC sector
and the other one produced by a Topcolor sector. We focus on the
phenomenological differences between the light composite Higgs present
in our model and the fundamental Higgs boson predicted by the Standard Model and
their production at the LHC. We show that in this scheme the main
production channel of the lighter Higgs boson are the associated
production with a gauge boson and $WW$ fusion but not the gluon-gluon
fusion channel which is substantially suppressed.  
\end{abstract}

\section{Introduction}

High Energy Physics is at the dawn of a new era. In the very near
future the LHC will allow us to explore, for the first time, the physics
responsible for the electroweak symmetry breaking. We are quite sure
that something different to the minimal Higgs sector of the Standard
Model will be found, but the conjectured alternatives, far from being
unique, are varied and very different among them, going from Higgsless
models to Supersymmetry.

A still very appealing possibility is the existence of a new strong
interaction at the TeV scale that breaks the electroweak symmetry
by the formation of condensates of (techni-)fermions. This kind of
models receives the generic name of Technicolor (TC) \cite{Hill:2002ap}.
Nevertheless, the idea of a new strong dynamics responsible for the
mass of the gauge bosons has been seriously challenged by the precision
measurements made at LEP. The main source for the discrepancy was
the underlying and unnecessary hypothesis that Technicolor was essentially
a scaled up version of QCD. Nevertheless, during the last few years,
new models of Technicolor have been built exhibiting interesting properties
like: minimal number of technifermions leading to acceptable values
of the $S$ parameter, quasi-conformal behavior of the coupling constant
({}``walking'' Technicolor) and the presence, in the low energy
spectrum, of a light Higgs-like composite scalar
\cite{Dietrich:2005jn,Dietrich:2005wk,Sannino:2005dy,Gudnason:2006ug,Foadi:2007ue,Ryttov:2008xe}.  
Phenomenological studies have been done, showing that such a composite
Higgs boson and the composite vector bosons predicted by the models
are able to produce interesting signals at the LHC such as the
enhancement, with respect to the Standard Model, of the associated
production of a Higgs and a gauge boson 
\cite{Zerwekh:2005wh,Belyaev:2008yj}. 

Despite the success of these new ideas, a fundamental problem remains
in such a dynamical approach to the electroweak physics: the origin
of the fermion masses. A traditional solution has been to extend the
Technicolor interaction to include standard fermions. This idea works
very well for the two lighter generations, but is unable to explain
the huge mass of the top quark. A viable solution is the so called
Topcolor Assisted Technicolor\cite{Hill:1994hp} where a new strong
interaction,Topcolor (TopC) generates a dynamical mass for the top
quark while Technicolor provides mass to the $W^{\pm}$ and $Z^{0}$gauge
bosons and the masses of the fermions belonging to the first two families
are generated by Extended Technicolor (ETC) . 

In this paper, we take seriously these ideas and we propose that the
low energy limit of an underlying Topcolor Assisted Technicolor theory
must contain in its spectrum two composite Higgs doublets: one for
generating the masses of the the $W^{\pm}$ and $Z^{0}$ gauge bosons
and the masses of the light fermions and another for generating the
masses of the fermions of the third generation. We emphasize that the
phenomenology of the resulting light Higgs boson is necessarily very
different from the standard expectations.

As an example, we construct an effective low energy Lagrangian that
combine the main features of (Minimal) Walking Technicolor and Topcolor
Assisted Technicolor. We assume, for example, that MWT is able to
produce a light composite Higgs boson that, from the effective theory
point of view, is able to break the electroweak symmetry and gives
mass (via Yukawa coupling that have their origin in ETC) to the first
two families of fermions. On the other hand, we postulate the existence
of an additional and maybe heavier composite Higgs boson coming from
a Topcolor sector that only couples to the third family. We include also
some vector resonances for each strong sector.

\section{The Model}

We start by generalizing the phenomenological model introduced in
reference \cite{Zerwekh:2005wh}. In the present description we are
going to include, beside the fields of the Standard Model (which we
call $V_0$ and $B_0$), two isotriplet
spin 1 fields (which we call $V_{1}$ and $V_{2}$), two isosinglet
spin 1 fields ($B_{1}$, $B_{2}$) and two complex scalar doublets:
$\varphi$ and $\Phi$. We assume that $V_{1}$, $B_{1}$ and $\varphi$
are composite fields produced by Technicolor and couple to the first
and second fermion generations, while $V_{2}$, $B_{2}$
and $\Phi$ are Topcolor bound states and couple to the third
generation. If we work in a basis where 
$V_{0}$ (the elementary {}``proto''-$W$ vector field), $V_{1}$
and $V_{2}$ transform as gauge fields under $SU(2)_{L}$ while $B_0$,
$B_1$ and $B_2$ transform as gauge fields of $U(1)$;  the Lagrangian
of the gauge sector can be written as:

\begin{eqnarray}
\mathcal{L}_{\mathrm{gauge}} & = & -\frac{1}{4}V_{0\mu\nu}^{a}V_{0}^{a\mu\nu}-\frac{1}{4}V_{1\mu\nu}^{a}V_{1}^{a\mu\nu}-\frac{1}{4}V_{2\mu\nu}^{a}V_{2}^{a\mu\nu}\nonumber \\
 &  & +\frac{M_{1}^{2}}{2g_{1}^{2}}(g_{0}V_{0}-g_{1}V_{1})^{2}+\frac{M_{2}^{2}}{2g_{2}^{2}}(g_{1}V_{1}-g_{2}V_{2})^{2}\nonumber \\
 &  & -\frac{1}{4}B_{0\mu\nu}B_{0}^{\mu\nu}-\frac{1}{4}B_{1\mu\nu}B_{1}^{\mu\nu}+\frac{M_{3}^{2}}{2g_{1}^{'2}}(g'_{0}B_{0}-g'_{1}B_{1})^{2}\nonumber \\
 &  &
 +\frac{M_{4}^{2}}{2g_{2}^{'2}}(g'_{1}B_{1}-g'_{2}B_{2})^{2}\label{eq:LagrGauge} 
\end{eqnarray} 
where $g_i$ ($g'_i$) are the coupling constants associated to the
vector bosons $V_i$($B_i$) (with $i=0,1,2$).

At this point, we want to point out that it would be straightforward
to introduce axial-vector fields which would be necessary to minimize
the value of the $S$ parameter at the phenomenological level, nevertheless
we include only vector resonances in order to focus on the main features
of the model, specially those related to the composite Higgs.

For the Higgs sector, we use a CP-conserving two doublets Higgs Lagrangian
\cite{Gunion:2002zf}:

\begin{eqnarray}
\mathcal{L}_{\mathrm{Higgs}} & = & (D_{\mu}\varphi)^{\dagger}(D^{\mu}\varphi)+(D_{\mu}\Phi)^{\dagger}(D^{\mu}\Phi)\nonumber \\
 &  & -\mu_{1}^{2}\left(\varphi^{\dagger}\varphi\right)-\mu_{2}^{2}\left(\Phi^{\dagger}\Phi\right)+\left[\mu_{12}^{2}\left(\Phi^{\dagger}\varphi\right)+\mathrm{h.c.}\right]\nonumber \\
 &  &
 -\frac{\lambda_{1}}{2}\left(\varphi^{\dagger}\varphi\right)^{2}-\frac{\lambda_{2}}{2}\left(\Phi^{\dagger}\Phi\right)^{2}-\lambda_{3}\left(\Phi^{\dagger}\Phi\right)\left(\varphi^{\dagger}\varphi\right)
\nonumber \\
 &  &
 -\lambda_{4}\left(\Phi^{\dagger}\varphi\right)\left(\varphi^{\dagger}\Phi\right)
-\left[\frac{1}{2}\lambda_{5}\left(\Phi^{\dagger}\varphi\right)^{2}
+\mathrm{h.c.}\right]
\label{eq:HiggsSector}
\end{eqnarray}

where

\begin{eqnarray}
D_{\mu}\varphi & =
& \partial_{\mu}\varphi+i\tau^{a}g_{0}\left(1-f\right)V_{0\mu}^{a}\varphi
\nonumber \\
 &  &
 +i\tau^{a}g_{1}fV_{1\mu}^{a}\varphi+i(1-k)\frac{Y}{2}g'_{0}B_{0\mu}\varphi
\nonumber \\
 &  & +ik\frac{Y}{2}g'_{1}B_{1\mu}\varphi
\label{eq:CovDerivHiggsLiv}
\end{eqnarray}
,

\begin{equation}
D_{\mu}\Phi=\left(\partial_{\mu}+i\tau^{a}g_{2}V_{2\mu}^{a}
+i\frac{Y}{2}g'_{2}B_{2\mu}\right)\Phi 
\label{eq:CovDerivHiggsPes}
\end{equation}
 and $f$ and $k$ are phenomenological parameters that assure gauge
invariance.

%************** Addition Revision 2************************ 
We want now to emphasize an important aspect of our
model. Because we are working in a basis where $V_0$, $V_1$ and $V_2$
(all of them) transform  as gauge fields of the \emph{same group} $SU(2)_L$,
any linear combination $\alpha_0 g_0 V_0 + \alpha_1 g_1 V_1 +
\alpha_2 g_2 V_2$ ( where $\alpha_0$, $\alpha_1$ and $\alpha_2$ are
some parameters) transforms like a gauge field and can be used to
construct a covariant derivative, provided that
$\alpha_0+\alpha_1+\alpha_2=1$. Evidently, a similar feature happens
in the $U(1)_Y$ sector. This property allows us to couple $\Phi$ only
with $V_2$ and $B_2$ as in equation (\ref{eq:CovDerivHiggsPes}) and,
therefore,suppress the contribution of $\Phi$ to the mass of the
$W$ and $Z$ bosons because any contribution would come only from a
mixing term (assumed to be small). On the other hand, in order to be
consistent with the known phenomenology, we can always include in the
covariant derivative, in the Dirac term of the third generation
quarks, the general linear combination of vector fields in order to
guarantee that the top quark couples to the $W$ and $Z$ bosons in the
usual way. More details about this method of treating vector
resonances can be found in \cite{Zerwekh:2003zz} and
\cite{Zerwekh:2005wh}.       
%**********************************************************

We assume that both scalar doublets get a vacuum expectation value:

\begin{equation}
\left\langle \varphi\right\rangle
=\frac{1}{\sqrt{2}}\left( \begin{array}{c}0\\v_{1}\end{array}\right),\;\left\langle
    \Phi\right\rangle=\frac{1}{\sqrt{2}}
\left( \begin{array}{c}0\\v_{2}\end{array}\right)
.
\label{eq:VEV}
\end{equation}

In order to simplify our model, focusing on the main physical ideas,
we are going to work in the limit where $g_{2}$ and $g'_{2}$ are
large enough to produce an effective decoupling of $V_{2\mu}^{a}$
and $B_{2\mu}$ from the rest of vector fields in (\ref{eq:LagrGauge}).
In this limit we re-obtain the model described in \cite{Zerwekh:2005wh}
for the TC-ETC sector plus an additional Topcolor sector. When the
Higgs doublets get their vacuum expectation values, the photon and
the $W^{\pm}$and $Z$ bosons can be written in terms of the original
fields as: 

\begin{eqnarray*}
A & = &
\frac{g'_{0}}{\sqrt{g_{0}^{2}+g{}_{0}^{'2}}}V_{0}^{3}+\frac{g_{0}}{\sqrt{g_{0}^{2}
+g{}_{0}^{'2}}}B_{0}
+\frac{g_{0}g'_{0}}{g_{1}\sqrt{g_{0}^{2}+g{}_{0}^{'2}}}V_{1}^{3}\\  
 &  & +\frac{g_{0}g'_{0}}{g_{1}\sqrt{g_{0}^{2}+g{}_{0}^{'2}}}B_{1}\\
Z & = &
\frac{g_{0}}{\sqrt{g_{0}^{2}+g{}_{0}^{'2}}}V_{0}^{3}-\frac{g'_{0}}{\sqrt{g_{0}^{2}
+g_{0}^{'2}}}B_{0}+\frac{g_{0}^{2}}{g_{1}\sqrt{g_{0}^{2}
+g_{0}^{2}}}V_{1}^{3}\\    
 &  & -\frac{g{}_{0}^{'2}}{g_{1}\sqrt{g_{0}^{2}+g{}_{0}^{'2}}}B_{1}\\
W^{\pm} & = & \frac{1}{\sqrt{2}}\left\{ \left(V_{0}^{1}\mp
    iV_{0}^{2}\right)+\frac{g_{0}}{g_{1}}\left(V_{1}^{1}\mp
    iV_{1}^{2}\right)\right\}
 \end{eqnarray*}

From (\ref{eq:CovDerivHiggsLiv}) and (\ref{eq:CovDerivHiggsPes})
we see that $\varphi$ couples to $V_{0},\: V_{1},\: B_{0}$ and $B_{1}$
while $\Phi$ only couples to the Topcolor generated vector resonances
($V_{2}$ and $B_{2}$). As a consequence, only $\varphi$ generates
the mass of the $W^{\pm}$. This fact obligates us to identify $v_{1}$with
the electroweak scale $246$ GeV while $v_{2}$ remains unconstrained.
Nevertheless, we still can suppose that $v_{2}$ can be of the order of
the weak scale and, hence, $v_{2}\sim v_{1}$.

%************** Correction number 1
We will use the vacuum expectation value of $\Phi$ for producing
the mass of the third generation fermions through an usual Yukawa
interaction.  Moreover, $\Phi$ generates
(additional) mass for the TopC vector resonances. As is usual in
models with two Higgs doublets, after the electroweak symmetry breaking
and diagonalizing the mass matrix of the Higgs sector, charged Higgs
bosons ($H^{\pm}$) and a CP-odd scalar ($A^0$) will appear in the
physical spectrum. In our scheme, they have their origin in $\Phi$ and
they correspond to the so called ``top-pion'' in Topcolor Assisted Technicolor
models. Therefore, they will couple strongly to the third
generation. On the other hand, we expect that
the lighter mass eigenstate, which will be mainly Technicolor-generated,
will couple to the third family fermions with a reduced strength (compared
to the SM).
%******************

%***************** Correction number 2
Let's continue with the study of the Higgs sector. Lagrangian
(\ref{eq:HiggsSector}) has many free parameters and we need to make
some simplifying assumptions. We chose to set up the following
relations: $\mu_{12}=0$, $\lambda_4=\lambda_5$ and $\lambda_5<0$. This
set of conditions keeps our expressions simple and still leads to an
acceptable phenomenology. For example, $\mu_{12}=0$ is well motivated
because it helps to restrict
possible FCNC processes at tree and one loop level
\cite{Gunion:2002zf}. With this choice, the minimum of the potential
is obtained for: 
%***************************************

\begin{eqnarray}
\mu_{1}^{2} & = &
-\frac{1}{2}\lambda_{1}v_{1}^{2}-\frac{1}{2}\lambda_{3}v_{2}^{2} \nonumber \\
\mu_{2}^{2} & = &
-\frac{1}{2}\lambda_{2}v_{2}^{2}-\frac{1}{2}\lambda_{3}v_{1}^{2}
\label{eq:HiggsMinimum} 
\end{eqnarray}

Of course, in order to have a true minimum we need that $\lambda_{1}>0$,
$\lambda_{2}>0$ and $\lambda_{3}>-\sqrt{\lambda_{1}\lambda_{2}}$.

%********** change number 3
The masses of the charged Higgs and the CP-odd scalar are:

\begin{equation}
  \label{massA}
  m^2_{H^{\pm}}=m^2_{A^0}=-v^2\lambda_5
\end{equation}
\noindent
where $v^2=v^2_1+v_2^2$. We want to emphasize again that $v_2$ is not
fixed by the mass of the $W$ and, in principle, can be large enough to
maintain $ m^2_{H^{\pm}}$ and $m^2_{A^0}$ above their experimental limit
while keeping $\lambda_5$ relatively small. In fact, because
$\lambda_5$ and $\lambda_3$ are produced by the same underlying
dynamics, it is not unreasonable to expect that $|\lambda_5|$ and
$|\lambda_3|$ must be of the same order of magnitude.

 In general, it is expected
that top-pion's mass is of the order $200-400$
GeV \cite{Hill:2002ap}.
In our case, we can apply the limits on $m_{H^{\pm}}$ that have been
obtained for the Two Higgs Doublet Model type III. In this context, it
has been recently shown that FCNC restrictions imply that $m_{H^{\pm}}
> 300$ GeV \cite{Mahmoudi:2009zx}.   
%******************************************** 

Let's denote by $\tilde{h}$ and $\tilde{H}$ the radial excitations
of $\varphi$ and $\Phi$ respectively around the new vacuum. The
Higgs mass matrix in the basis $\left(\tilde{h},\tilde{H}\right)$
is:

\begin{equation}
\mathcal{M}^{2}=\left(\begin{array}{cc}
\lambda_{1}v_{1}^{2} & (\lambda_{3}+2\lambda_5) v_{1}v_{2}\\
(\lambda_{3}+2\lambda_5) v_{1}v_{2} &
\lambda_{2}v_{2}^{2}\end{array}\right)\label{eq:HiggsMassMatrix}
\end{equation}

In the limit where  $\lambda_{3}+2\lambda_5$  is small, we can write
the eigenvectors as:

\begin{eqnarray}
h & = & \tilde{h}-\theta\tilde{H}\nonumber \\
H & = & \tilde{H}+\theta\tilde{h}\label{eq:HiggsEigenvectors}
\end{eqnarray}
\noindent
where $\theta$ is the mixing angle and is given by the following
expression:

\begin{equation}
\theta=\frac{v_{1}v_{2}(\lambda_{3}+2\lambda_5)}{\left(m_{H}^{2}-m_{h}^{2}\right)}. 
\end{equation}

In the expression above, $m_{h}$ and $m_{H}$ are the masses of $h$ and
$H$ respectively 
and, at this level of approximation, they are given by:

\begin{eqnarray}
m_{h}^{2} & = & \frac{1}{2}\lambda_{1}v_{1}^{2}\nonumber \\
m_{H}^{2} & = &
\frac{1}{2}\lambda_{2}v_{2}^{2}\label{eq:HiggsEigenvalues}
\end{eqnarray}

At this point, it is worth to be remembered that $h$ is a mainly
TC generated particle while $H$ is mainly a Topcolor composite state
and we have assume from the beginning that $H$ is considerable heavier
than $h$.

\section{Results}

One of the main consequences of the picture described above is that
the light Higgs boson $h$ couples weakly to the top quark. By hypotheses,
only $\tilde{H}$ couples to the third generation of fermions and,
hence, $h$ gets its coupling to the top by the mixing term. Consequently,
the coupling is proportional to $\theta$ which is assumed to be small.
This means, also, that the effective $h$-gluon-gluon interaction
(which is responsible for the main production mechanism of the standard
Higgs at the LHC) will be significantly suppressed. Additionally,
the decay of $h$ to $b\bar{b}$ will be also suppressed by the same
reasons.

On the other hand, $h$ couples in the standard way to the gauge bosons
$W^{\pm}$ and $Z$; but also couples to Technicolor vector resonances.
This fact implies that its associated production with gauge bosons
will be enhanced as was found in \cite{Zerwekh:2005wh} and \cite{Belyaev:2008yj}.

By contrast, $H$ couples in the normal way to the top quark. Additionally
, in our simple model where we have completely decoupled the Topcolor
generated vector resonances $V_{2}$ and $B_{2}$ from the other vector
bosons, $\Phi$ (and hence $\tilde{H}$) doesn't couple to the gauge
bosons $W^{\pm}$ and $Z$ (see equation (\ref{eq:CovDerivHiggsPes})).
That means that $H$ couples weakly to the $W^{\pm}$ and $Z$, through
it $\tilde{h}$ component. The strength of this interaction will be
proportional to the small mixing angle $\theta$.

\section{Conclusions}

In this work, we have proposed a scenario with two composite Higgs
doublets that can be a viable low energy limit of an underlying
complete and natural 
strong electroweak symmetry breaking sector which aims to explain not
only the mass of the gauge bosons, but also
the mass of the fermions. As a proof of concept,
we constructed a model which implements this idea. It includes a Technicolor
generated Higgs doublet which breaks the electroweak symmetry and
gives mass to the first two generation of fermions, while a second,
Topcolor generated, Higgs doublet gives mass only to the third generation
fermions. An important ingredient of the model is the inclusion of
vector resonances which allows us to implement the role separation
between the two Higgs doublets. The different roles played by
the two Higgs doublets implies that the phenomenology of the light
composite Higgs boson is expected to be very different from that of
the standard Higgs. In particular, we expect that its main production
mechanisms at the LHC were the associated production with gauge bosons and $WW$
fusion, but not gluon fusion as happens in the Standard Model.
\section*{Acknowledgments}
I want to thank to Rogério Rosenfeld and Claudio Dib for very interesting
discussions in the early stages of this work. I am also in debt with
Maximiliano Rivera for the useful discussions about the potential
of a two Higgs doublet model. The author is partially supported by
Fondecyt grant 1070880 and by the Center of Subatomic Studies. TGD. 

% \bibliographystyle{h-physrev}
% \bibliography{biblio}

\end{document}